# Title

Bio-inspired method based on bone architecture to optimize the structure of mechanical workspieces

# Auteurs

Clément Audibert[a], Julien Chaves-Jacob [a,1], Jean-Marc Linares [a], Quentin-Alexis Lopez [a]

[a] *Aix Marseille Univ, CNRS, ISM, Marseille, France*

# Abstract

Nowadays, additive manufacturing processes greatly simplify the production of openwork workpiece providing new opportunities for workpieces design. Based on Nature knowledge, a new bio-inspired workpiece structural optimization approach is presented in this paper. This approach is derived from bones structure. The aim of this method is to reduce the workpiece weight maintaining an acceptable resistance. Like in bones, the porosity of the part to optimize was controlled by a bio-inspired method as function of the local stress field. Shape, size and orientation of the porosities were derived from bone structure; two main strategies were used: one inspired of avian species and other inspired of terrestrial mammalian. Subsequently, to validate this method, an experimental test was carried out for comparing a topological optimization and the proposed bio-inspired designs. This test was conducted on a beam part in 2.5D subjected to a static three-point bending with 65% of density. Three beams were manufactured by 3D metal printing: two bio-inspired beams (terrestrial mammalian and avian species) and the last designed using a topological optimization method. Experimental test results demonstrated the usefulness of the proposed method. This bio-inspired structural optimization approach opens up new prospects in design of openwork workpiece.

# Keywords

Bio-inspired optimization; Functionally graded materials; Topological optimization; 3D printing; Bone structure

# Nomenclature

$p$ pattern size
$c$ porosity mean radius
$e$ minimal distance between two porosities
$a$ Principal ellipse axis
$b$ Second ellipse axis
$h$ characteristic structure length
$E$ Young modulus
$\nu$ Poisson ratio
$\rho_f$ Volume fraction
$\sigma_1$ Principal tensile stress
$\sigma_3$ Principal compression stress

## 1 Introduction

For each lifestyle, nature takes into account the constraints imposed by the environment to create optimized biological systems in term of mechanical resistance, mass and lifespan. Then, porous materials appeared in lots of systems in nature. Bone structure is the result of millions of years of evolution. Thus, the bone is still considered as one of the most efficiency structure architecture (cortical and trabecular). It has been the subject of many research studies to understand the driving parameters of its micro-structure. Bone cells are able to measure the local stress acting as a bio-mechanical sensor. Bone material structure and its distribution are adapted by collaborative work of osteocyte and osteoblast cells versus the mechanical load [1]. Trabecular bone micro-structure aligned in the principal stress directions (Wolff's law)[2] is generated by this natural adaptation (Fig .1). Different volume fractions are induced by the local stress level. The obtained result is a combination of plate-like elements and rode-like elements [3]. Finite element analysis performed on the

---

trabecular bone indicate that this material distribution gives better resistance to compression and shear loading [4]. Research works [5] have demonstrated that the porosity shape can be well fitted by an ellipsoid where their eigen directions are aligned with the principal stress direction. A better compromise between lightweight, high stiffness, damage-tolerance and robustness with respect to multiple loadings [6] is obtained by this natural optimization.

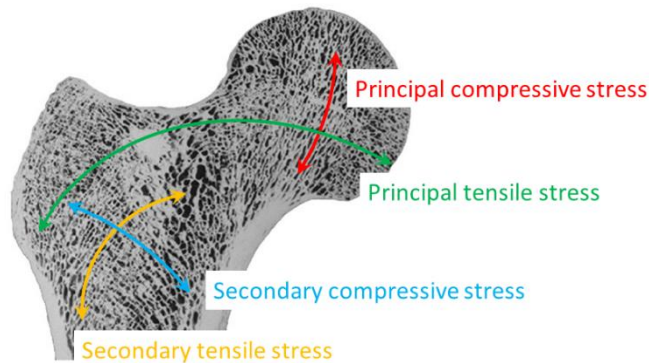

**Fig. 1.** The basis of Wolff's trajectorial theory: A midfrontal section of the proximal femur, showing trabecular architecture with schematic diagram of the important anisotropic loading groups (Adapt from[7]).

Bone structure has inspired many works to obtain light structures with high mechanical properties. These studies have conducted to several algorithms of topology optimization. *Daynes et al.*[8] developed a functionally graded lattice core structure and bio-inspired on the trabecular bone. Thanks to an optimization process, the authors created beam elements along the principal stress directions of a workpiece subjected to three-point bending. The optimized beam design has demonstrated a stiffness and a strength improvement respectively of 101% and 172% compared to a standard lattice structure [8]. This design was composed only by beam elements whereas the trabecular bone is composed of plate elements and beam elements in order to obtain a better stress distribution. Thus, probably, discontinuity and stress concentration could be introduced by beam elements. This could lead to a low damage tolerance, and a low fatigue life.

In the other hand, *Ying et al.*[9] presented a method based on anisotropic centroidal Voronoi tessellations and an iteratively optimization framework. They obtained a 2.5D anisotropic porous structure with a strength increase of 34% comparing to an isotropic porous structure. Thus, a better material continuity and a better stress distribution can be expected. These anisotropic porosities are similar with the trabeculae shape identified in the literature. However, the authors have highlighted that the extension of 2.5D to 3D shapes could induced a high computational cost and a high geometric complexity of the 3D anisotropic Voronoi cells.

The results of a topological optimization depend of the initial structure and the boundary conditions. In the case of many mesh elements, these numerical results are dependent of iterations involving a high computation time. These models converge to a structure that minimizes strain energy and maximizes rigidity. However, they do not take into account fatigue durability, resistance to dynamic stress, and load variations. Nature, after billions of years of evolution has generated biologic structures. These structures are a good compromise for satisfying all the various boundary conditions imposed by its environment. Loading variations are withstood by these biological structures, making them mechanical reliable.

In this paper, a new bio-inspired method is proposed to obtain porous structures. The functionally graded porous structure is bio-inspired on the bone while the skin of the workpiece is dense (as in bone, cortical bone). The proposed method do not use of an iterative optimization strategy. Shape, size and orientation of the porosities were derived of bones structure; two main strategies are use: one inspired of avian species and other inspired of terrestrial mammalian. Thereafter, to validate the usefulness of the proposed method a comparison was carried out between the bio-inspired beams and a beam deriving from a topological optimization method. The experimental validation was conducted with a beam part in 2.5D (with 65% of material density) subjected to a static three-point bending. Three beam were additively manufactured by Selective Laser Melting (SLM) with 17-4-PH steel material: two beams obtained by bio-inspired designs (avian species and terrestrial mammalian) and the last obtained by topological optimization (based on the work of *Daynes et al.* [8]). A finite element analysis was performed on the three beams in order to evaluate the stress field in each beam and predict their behaviors at breaking. In order to estimate mechanical performances, a three-point bending test was performed for the three designs.

## 2 Materials and Methods

### 2.1 Bio-inspired optimization method

Existing topological optimization methods are based on the minimization of the elastic strain energy. Hence, considering a given stress field, a material distribution is generated by the algorithm to maximize the structure compliance. Structures are optimized without considering other parameters as fatigue life or local contact pressure. To overcome these difficulties, the proposed approach in this paper is based on the billions of years of the nature skill evolution [10], [11]. Indeed, the trabecular architecture and the cortical bone makes possible to achieve a strong compromise between stiffness, weight, fatigue resistance, impact resistance. The local strain at the trabeculae connections enables to ensure a better distribution of the contact pressure [12], [13]. This structure meets a variety of specifications and have been adapted to the lifestyle of each living species [14], [15], [16].

In this context, the correlation between bone structure of wild animal species and their environment loading have led to the bio-inspired algorithm proposed in this paper. In consequence, this method implicitly integrates all the optimization parameters considering by nature. As shown in Fig. 2, this algorithm comprises several stages. The first stage is the estimation of the workpiece's stress field, considered fully dense. Then, a bio-inspired discretization according to the structure size is realized. For each discretized element, a volume fraction is calculated as a function of local stress field. It allows generating a network of ellipsoid porosities oriented in the principal axes of stress similarly to the bone observations. In this paper, the obtained porous structures were manufactured by 3D metal printing using a powder nest technology.

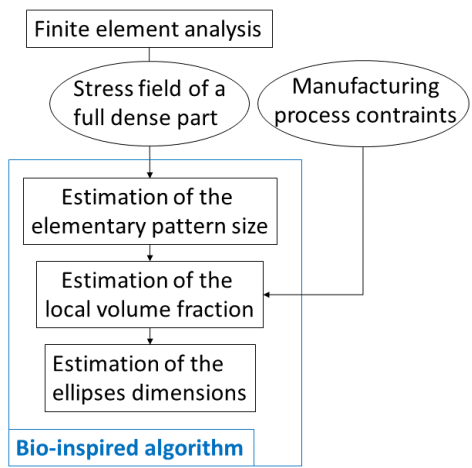

**Fig. 2.** Bio-inspired algorithm

### 2.1.1 Bio-inspired-elementary pattern

Trabecular bone is composed of elliptical porosities with a preferential orientation in the main stress directions [7], [17]. As part of the creation of a structural optimization algorithm, an elementary bio-inspired pattern is proposed. Then, the elementary pattern proposed is a 2D square composed of a central ellipse whose axis **a** and **b** are oriented in the main directions of the stress tensor. The dimensions ‖**a**‖ and ‖**b**‖are function of the local stress level (see Fig. 3.(a)). To obtain a higher porosity fraction, and reduce the workpiece weight, four additional ellipses are arranged at each corner of the mesh (see Fig. 3.(b)). For ensuring a continuous evolution of the porosity gradient in the structure, these peripheral ellipses are calculated as the result of the average of the four ellipses that surround it.

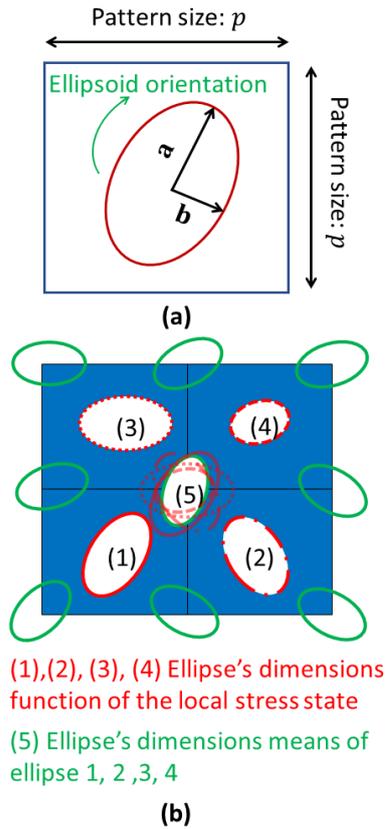

**Fig. 3.** (a) Elementary pattern with a central ellipsoid; (b) Ellipsoid building schema

*2.1.2  Bio-inspired pattern size*

In order to optimize the workpiece, the initial CAD model is discretized using quadrangles with a length parameter $p$. The determination of this parameter is based on the observations of Doube *et al.* [18]. The authors observed that the trabecular bone structure is dependent of the mass and the size of the animal. They demonstrated that the trabecular parameters (spacing, thickness and connectivity density) were power function of the femoral head diameter which was proportional to the animal size. Moreover, they highlighted a difference between terrestrial mammalian and avian species. It highlights the influence of the lifestyle on the trabecular structuration. Similarly, Ryan and Shaw [19] have shown, on primate bones, that this model applies as well to the femur as to the humerus. Thus, the porosity size is linked to the animal's dimension, but this size seems to be independent of the type of bone.

Based on these works, the pattern size was set as a power function of a characteristic length which was the femoral head diameter. Thus, this characteristic length can be identified on a structure as the highest distance from the neutral fiber of the workpiece. For example, in the case of a mechanical part, like a ring spanner, this characteristic length was the half height of the higher stress area (Fig. 4). Since the porosities of each pattern were oriented in the principal stress directions, the discretization direction does not affect the result of the optimization process.

In the present study, two discretization parameters have been used: one inspired on terrestrial mammalian (Eq(1)) and the other one inspired on avian species (Eq(2)) [18]. Thus, the bio-inspired discretization allows taking into account the scale effect.

terrestrial mammalian: $p = \left(\frac{h}{2}\right)^{0.407}$  (1)

Avian species: $p = \left(\frac{h}{2}\right)^{0.853}$  (2)

With,

$p$: pattern size

$h$: characteristic length of the structure

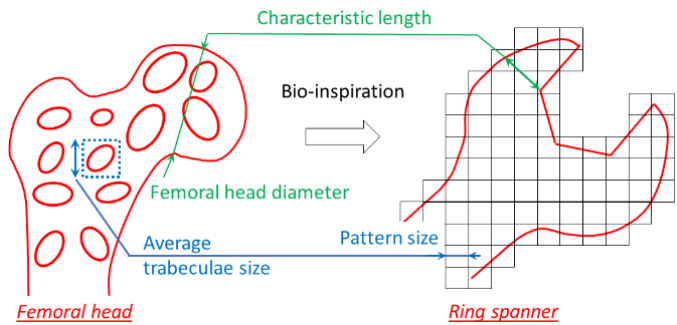

**Fig. 4.** Schema of bio-inspired discretization, example of application on a mechanical workpiece (example on a ring spanner)

### 2.1.3 Local bio-inspired estimation of the pattern material volume fraction

The natural optimization of the trabecular bone can be explained by the ability of the bone to create a link between the material volume fraction and the local stress field. This function is performed by osteocyte cells acting as a stress-sensor which conduct to a local adaptation of the porosity network [2].

The relation between the stress field and $\rho_f$ was bio-inspired on the trabecular femoral structure. The multiaxial loading to which the human femoral bone was subject, conducts to a structure optimized in compression, bending and tensile. It makes the femur a good candidate to identify a local volume fraction.

Next, the objective was to characterize the law between the local stress field and the volume fraction of material in a human femur. To do this, the stress field was initially estimated from a 2D finite element analysis (FEA) of a full dense homogenous femur having the elastic mechanical properties of the cortical bone ($E = 20$ GPa and $\nu = 0.3$ [20]). The considered load corresponds to the passage on one leg when walking [21]. Model boundary conditions and the computed Von Mises stress distribution are presented in Fig. 5.(a). Then, $\rho_f$ was determined by correspondence between the local microstructures (identified in the literature [22]) and the local stress field of a dense femur. The local volume fraction data was plotted on Fig. 5.(b) according to the Von Mises stress normalized by the elastic threshold of the cortical bone. The curve, which follows a power law (Eq. (3)), provides a fine determination coefficient ($R^2$ =0.97). The observations made by Gibson et al. [23] on trabecular bone behavior in compression give similar results. It appears that the trabecular bone becomes denser until it reaches 15% of the elastic limit of the cortical bone. From this stress field value, the bone structure becomes fully dense, named cortical bone.

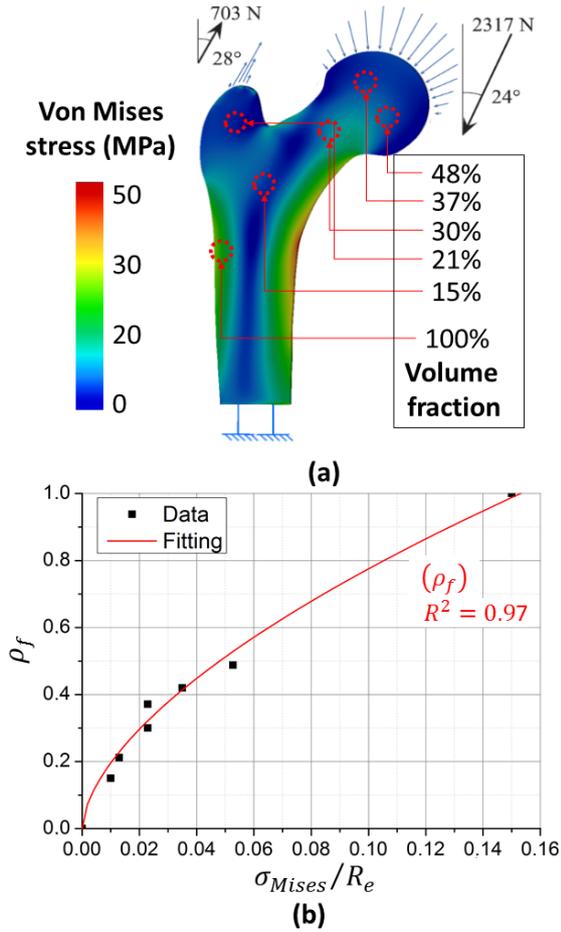

**Fig. 5.** (a) A human femur under a mechanical loading corresponding to the loading of a leg during walking [21]: Von Mises stress distribution and local volume fraction; (b) Graph of the local volume fraction versus local stress field

$$\rho_f = k_1 \left(\frac{S_{Mises}}{Re}\right)^{k_2} \quad (3)$$

With,

$\rho_f$ : Volume fraction

$S_{Mises}$ : Von Mises stress

$Re$ : 200 MPa, elastic threshold of the cortical bone

$k_1 = 3.07$ and $k_2 = 0.59$ : Function parameters

In a 2D case, to avoid the loss of material cohesion, a maximum porosity level has been set equal to 50 %. Indeed, regarding the porous material, the Hill model [26] has demonstrated the existence of a percolation threshold equals to 50 %. Beyond this level, the porosities collapse leading to a dramatically decreased of mechanical properties. This phenomenon is illustrated in Fig. 6. with 30 % of porosity fraction the spatial distribution of the holes creates a material network able to distribute mechanical stress. With 50 % of porosity fraction, the wall between holes becomes extremely thin. Beyond this point, with 70% of porosity for example, the material cohesion is lost, and it cannot distribute the mechanical stress anymore.

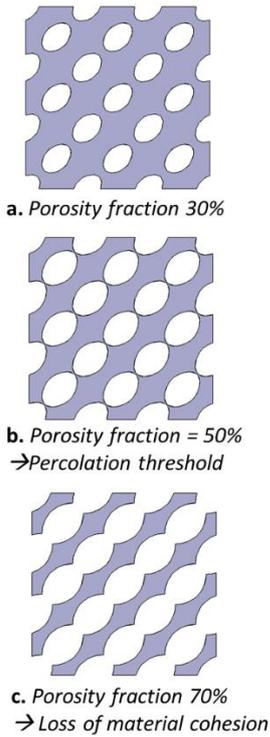

**a.** Porosity fraction 30%

**b.** Porosity fraction = 50%
→Percolation threshold

**c.** Porosity fraction 70%
→ Loss of material cohesion

**Fig. 6.** Material cohesion function of porosity ratio: a. Porosity fraction =30%; b. Porosity fraction = 50%; c. Porosity fraction = 70%

### 2.1.4 Bio-inspired estimation of the ellipses dimensions

Trabecular porosities are function of the bone region [3]. In the epiphysis region of a femoral bone, the trabecular pattern is a plate like element whereas in the metaphysis and the diaphysis regions there are rode elements [24], [25]. Some authors [20], [21], have quantified the microstructural anisotropy as a function of the main stress anisotropy (Fig. 7.).

After the identification of the volume fraction for each element, it was necessary to determine the ellipse dimensions and orientation. The literature data allows to link the ellipse form factor $FA$ to the main stress ratio $\left|\frac{\sigma_1}{\sigma_3}\right|$ by means of the exponential relation presented in Eq. (4). The obtained determination coefficient ($R^2$ =0.53) is in the same order of magnitude than literature values. This value was due to the high variability of the trabeculae. The ellipse form factor oscillates between 1 and 1.6. Thus, for a factor equals to 1, the porosity is a circle whereas for a factor equals to 1.6, the porosity is an ellipse with a major axis $\|\mathbf{a}\|$ 1.6 times larger than its minor axis $\|\mathbf{b}\|$.

$$FA = \frac{\|\mathbf{a}\|}{\|\mathbf{b}\|} = y_0 + C\, exp^{-\frac{1}{t}\left|\frac{\sigma_1}{\sigma_3}\right|} \quad (4)$$

With $\sigma_1 > \sigma_3$ : principal stresses
$y_0 = 1.6$
$t = 6.45$
$C = -0.6$
$\|\mathbf{a}\|$ and $\|\mathbf{b}\|$: dimensions of the principal ellipses axis

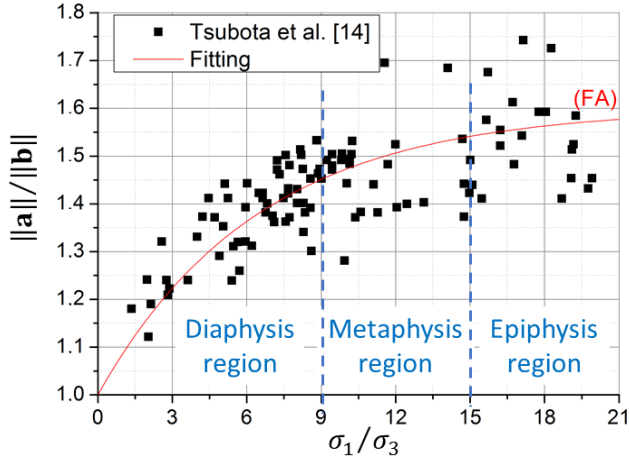

**Fig. 7.** Graph of ellipse form factor as exponential function of principal stress ratio. Based on literature data [21]

*2.2 Application test: three-point bending*

The algorithm was implemented through a VBA subroutine in CAD software. The bio-inspired algorithm was tested on a 2.5D beam in static three-point bending case. The beam dimensions were set as: $L = 150$ mm, $l = 15$ mm, $w = 30$ mm and the distance between bending supports was taken equals to 130 mm.

In section §2.1.3, a maximal threshold equals to 50 % of porosity was determined to ensure the continuity of the material. This porosity value may induce high stress concentration. A full analytical analysis of the effect of two porosities in an anisotropic medium was proposed by Berveiller *et al.* [27]. Indeed, to limit stress concentrations in a medium is necessary to ensure that the minimal distance between two porosities (noted *e*) is greater than the porosity size (Eq. (6)). This constraint will limit the maximal porosity rate. To simplify the calculation, the porosity shape was assumed cylindrical with a radius *c*. Its value was defined equal to the average of semi-minor and major ellipse axes size (Eq. (5)). Fig. 8. illustrates this assumption. Eq. (7) provides the volume fraction of this pattern. Eq. (8) shows the calculation of the minimum distance. Combining Eq. (6), (7) and (8) the minimal volume ratio was derived in (Eq. (9)). Its threshold value was determined at 65% to limit the stress concentration. Thereby, the experimental validation was carried out with this density value.

$$c = \frac{\|\mathbf{a}\| + \|\mathbf{b}\|}{2} \tag{5}$$

$$e \geq c \tag{6}$$

$$\rho_f = \frac{p^2 - 2\pi c^2}{p^2} \tag{7}$$

$$e = \frac{\sqrt{2}}{2} p - 2c \tag{8}$$

$$\rho_f \geq 1 - \frac{\pi}{9} \tag{9}$$
$$\rho_f \geq 65\%$$

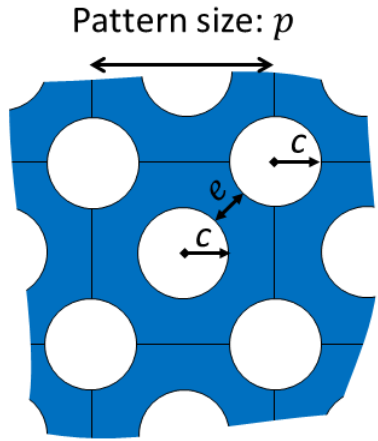

**Fig. 8.** Simplified model to determine the maximal porosity

*2.2.1   Optimized beam designs*

Two beam designs were proposed, one bio-inspired on terrestrial mammalian and another one bio-inspired on avian species.

First, a 2D finite element analysis was perform on a full dense beam solicited in three-point bending with a force of 5 kN. The stress field was used to determine the local volume fraction and orientation of ellipse axis for each element. Regarding the beam height, the element discretization was estimated to 3 mm in the case of terrestrial bio-inspired beam and 10 mm for avian bio-inspired beam. Next, the central ellipsoid of each pattern was generated, and then the peripheric ellipses were calculated as the mean of ellipses that surround it. To ensure its manufacturing, the minimal value of the semi minor ellipsoid axis was set equal to 0.55 mm. Under this limit it was impossible to extract the unmelt powder. The results at each steps of the optimization process are presented in Fig. 9.

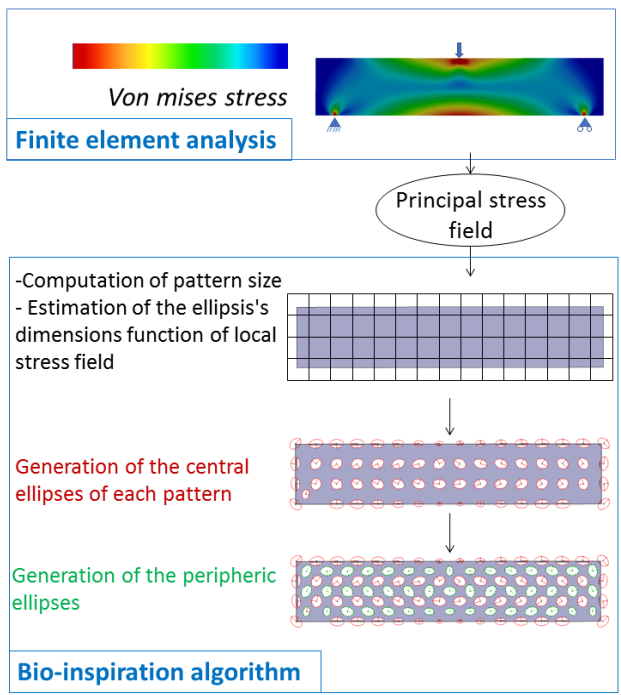

**Fig. 9.** Bio-inspired optimization process of a beam in three-point bending

In order to estimate the performance of bio-inspired beams, they were compared to a topological optimization design proposed by *Daynes et* al [8]. The authors applied their algorithm to a beam solicited in three-points bending using an optimization method based on isostatic line. The three designs tested in this paper are presented in Fig. 10.

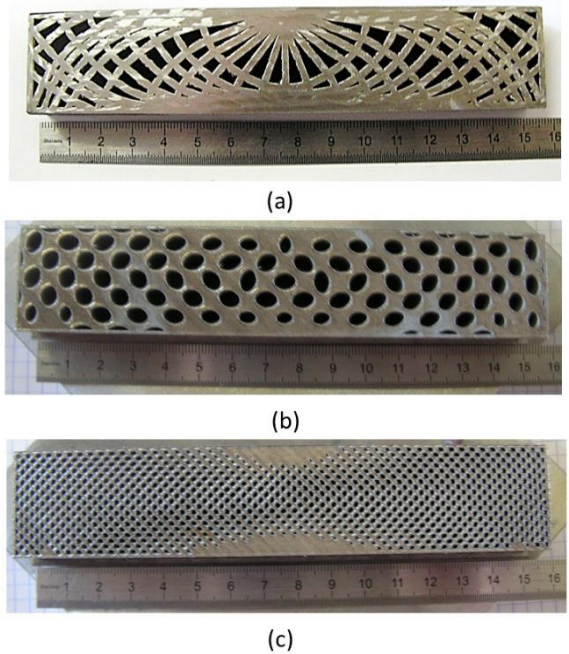

**Fig. 10.** Three-points bending designs: (a) Isostatic line optimized beam (adapt from [8]); (b) Avian bio-inspired beam; (c) Terrestrial bio-inspired beam

## 2.3 Materials and manufacturing process

The workpieces were additively manufactured by SLM technology with a Phenix PXM printer. The material used was 17-4 PH steel powder with a mean particle diameter of 26 μm. The printing parameters were chosen as follow: Laser power 189 watts, feed rate 1100 mm per second, gap between laser passes 65 micrometers, laser defocusing 4.6 mm (corresponding at a spot size around 0.1 mm) and layer thickness 30 micrometers. Once the workpieces built, compressed air was blown through the porous structure in a hermetic enclosure in order to remove unmelted powder. Then, the specimens were subjected to heat treatment at 1100°C during 1h followed by a water quenching. Finally, a last heat treatment was performed at 600°C during 4h followed by an ambient air-cooling. It eliminates the residual stress generated during the manufacturing process. The beams were finally removed from the built plate using a bandsaw and a milling operation was performed on the two faces to ensure a good parallelism of the faces.

### 2.3.1 Finite element analysis

A finite element analysis of the three-point bending test was performed on the three beam samples. The parts were meshed with 0.1 mm tetrahedral elements to ensure results convergence. The material law used was linear elastic with a Young's modulus of 205 GPa and a Poisson's ratio of 0.3. The bending supports were modeled as displacement boundary conditions, and an 80 kN static linear load was applied on the top of the beam.

### 2.3.2 Three-points bending test

Three-point bending tests were carried out on a Tema Concept machine under quasi-static conditions with a strain rate of 5 mm.min$^{-1}$ (Fig. 11). The punch head displacement and the force signal were recorded. To access to the strain field of the beam, the test was instrumented with a video extensometer using a digital image correlation system (DIC) provided by DANTEC.

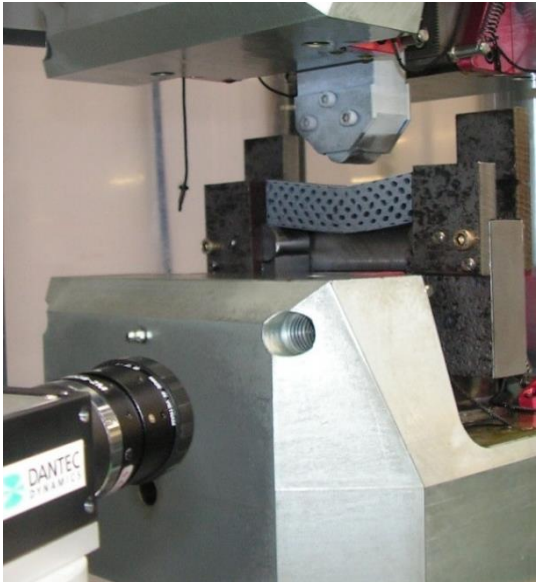

**Fig. 11.** Three-point bending test instrumented with an image correlation system

## 3 Results and discussion

### 3.1 FE analysis results: three-point bending

The Von Mises stress field of the three beams is presented in Fig. 12. Regarding the stress distribution, in all the three cases, the bending stresses are distributed on the upper part (compressive stress) and on the lower part (tensile stress). The terrestrial bio-inspired beam presents the best distribution of bending stresses. This may be explained by the small size of these porosities that allows a best stress repartition. On the other hand, the great porosity sizes of the isoline and the avian beams induce more stress concentration. Stress concentrations are also present at the fulcrum in the three cases. It is worth to note that, in the isoline case, because of the thin walls, the stress concentration in the sections could lead to a buckling failure mode. This nonlinear behavior cannot be predicted by a static simulation with an elastic-linear material model.

To conclude this section, the better stress distribution is in terrestrial Bio-inspired beam may suggest a best performance of this three designs.

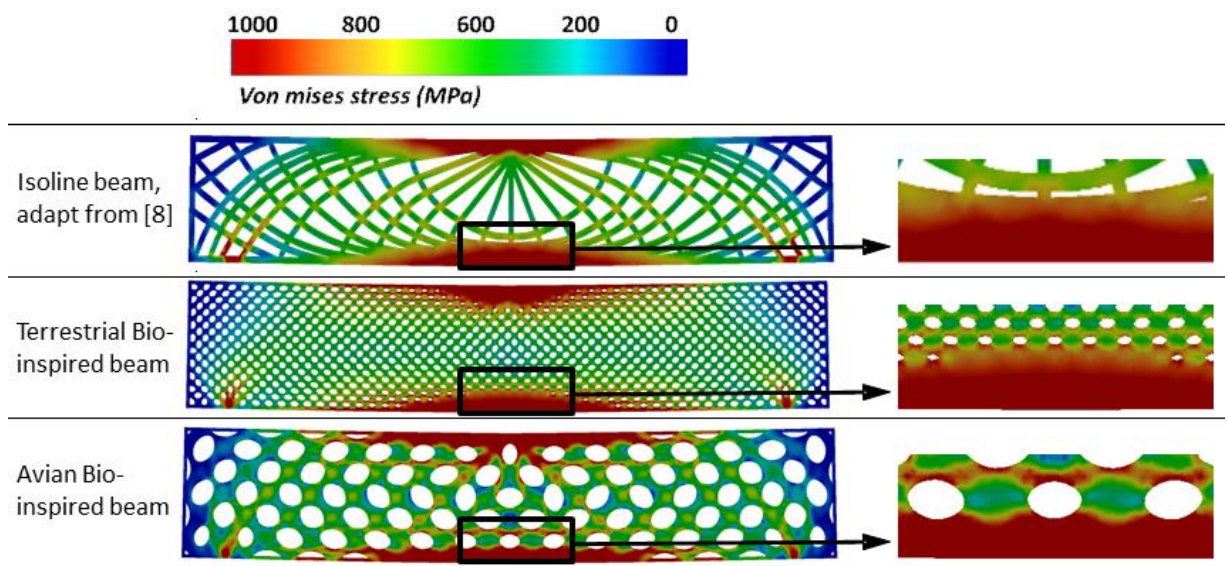

Fig. 12. Von Mises stress field in the three samples under three-point bending

## 3.2 Experimental test results

The experimental force-displacement curves of all three beams under three-point bending are shown in Fig. 13 (d). From 0 to 1.5 mm displacement, the three beams have demonstrated a linear behavior. Beyond this point, a nonlinear behavior can be observed followed by a load plateau. A shift of the initial stiffness and the stress plateau can be observed between the three beams.

These observations can be justified by the evolution of the strain field provided by the DIC system and presented in Fig. 14. These observations corroborate the FE analysis results. Initially, uniform strain field can be observed for all types of beam structures with a minimal through-thickness compression. A strain localization regarding to the three-point load application point was observed on the three beams (Fig. 14 (a), (b), (c)). Then the stress localization leads to the development of plastic strain.

The terrestrial bio-inspired beam has shown a failure initiation on the tensile side. It conducts to a shearing of nearest ellipses leading to the final failure (Fig. 14 (c)). The avian bio-inspired beam demonstrated also a similar failure occurring on the tensile side (Fig. 14 (b)).

On the other hand, the isoline sample demonstrated a progressive localization of plastic strain near the bending support (Fig. 14 (a)). It leads to the highly nonlinear behavior observed on the graph in Fig. 13 (d). This phenomenon explains the lower stiffness compared to terrestrial beam. It was followed by the plastic strain development which was responsible of the load plateau observed on the curve. Finally, the high compression strain near the force application point induced a strong plastic strain conducting to the final failure by buckling (Fig. 13 (a)).

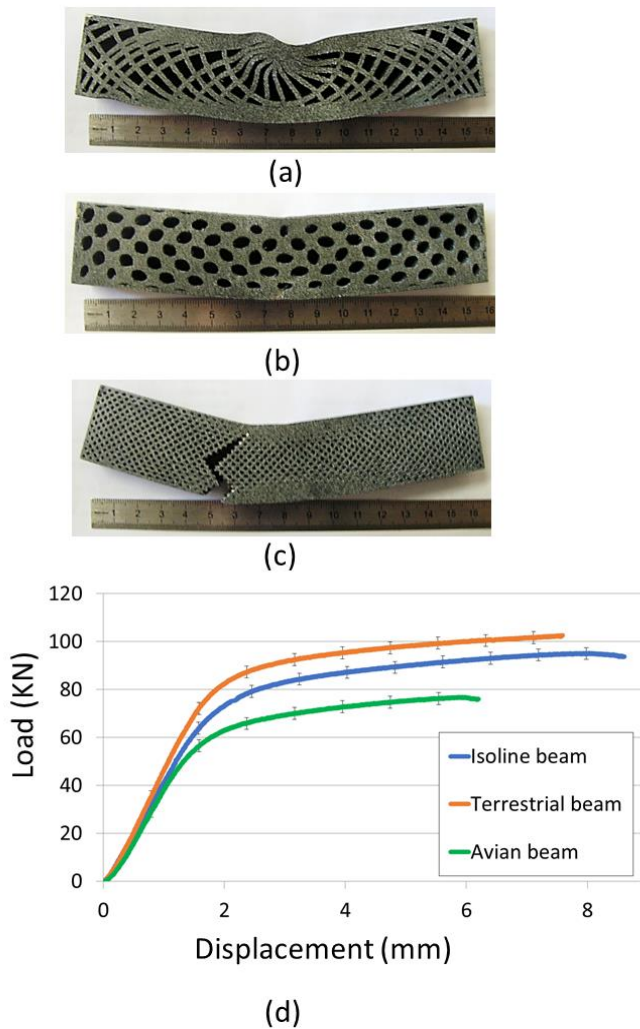

**Fig. 13**. Post-failure photography of: (a) Isostatic line optimized beam; (b) Avian bio-inspired beam; (c) Terrestrial bio-inspired beam. (d) Graph force-displacement with error bars of experimental bending tests

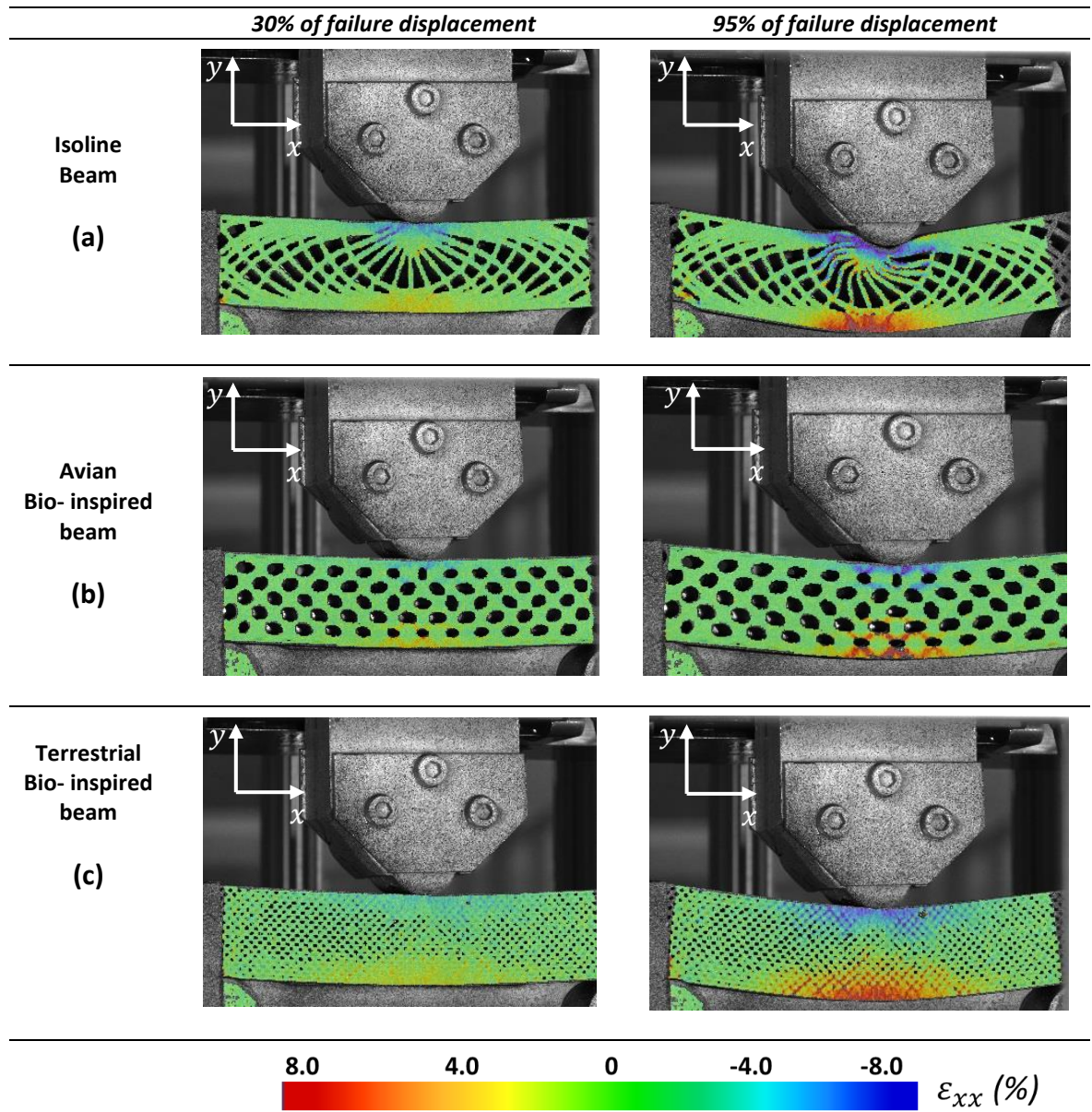

**Fig. 14**. Strain field evolution during bending test of: (a) Isostatic line optimized beam; (b) Avian bio-inspired beam; (c) Terrestrial bio-inspired beam

The comparison of the beams performance in term of stiffness and strength is presented in Table 1. The results show an increase by 17.34 % and 7.59 %, respectively for stiffness and strength for the terrestrial bio-beam compared to the isoline design. Whereas the avian beam demonstrates a stiffness and a strength, respectively lower by 1.27% and 19.3 %. These results can be explained by the DIC results, which evidences a strong localization of plastic strain for the isoline and avian beam. This contributes to lower mechanical properties.

The more uniform strain field until the sudden failure of the terrestrial beam have demonstrated the efficiency of the bio-algorithm to distributing stresses within the structure. Moreover, the higher stiffness of terrestrial beam compared to the avian bean could be explained by the link between the life style and the bone formation. Indeed, it was probable that the terrestrial mammalian bones were stiffer than avian species bones because of the higher pressure due to the higher weight which leads to a structure more efficient regarding the stress concentration.

**Table 1.** Beams performances in three-point bending test

|  | Isostatic line (adapt from [8]) | Terrestrial Bio-inspired beam | Avian Bio-inspired beam |
|---|---|---|---|
| Volume fraction (%) | 65.1±0.2 | 64.6±0.2 | 65.5±0.2 |
| Flexural stiffness (kN/mm) | 44.2±0.12 (reference) | 51.9±0.11 (+ 17.34 %) | 43.7±0.11 (-1.27 %) |
| Failure force (kN) | 95.2±2.5 (reference) | 102.6±2.5 (+7.59%) | 76.8±2.5 (-19.3 %) |
| Failure displacement (mm) | 8.6±0.01 (reference) | 7.5±0.01 (-11.8%) | 6.2±0.01 (-28.0 %) |
| Failure mode | Buckling | Traction | Traction |

*3.3 Discussion*

In the case of the topological optimization method, porosities were generated only considering strain minimization criteria in a purely elastic domain. However, during the test, the elastoplastic material behavior led to local plastic strain near the bending supports. These structural modifications changed the load strain distribution within the structure. Topological optimization did not consider load variations, which lead to premature failure by the local buckling structure. On the contrary, nature has generated structures that respond to multiple loads. The bio-inspired optimization method has provided geometries more resilient to changing loads. This has contributed to a better stresses distribution within the structure in the case of terrestrial bio inspired beam. It has limited the plastic strains under the supports conducting to a greater breaking load. These observations highlight the high dependence of the topological optimization to the initial boundary condition of the model. It could lead to workpiece very sensitive to stress localization. In the other hand the bio-inspired pattern seems to be more resilient to load concentration. The terrestrial bio-inspired design has demonstrated a better stress distribution than the topological design. The stress concentration has a great influence on the fatigue behavior. To confirm this hypothesis, it could be interesting to study the behavior of structural parts in other load cases. This future studies will allow to evaluate the fatigue life, damage tolerance and compare with the topological optimized workpiece performances.

The algorithm presented in this paper is part of a progressive approach, with a first step which enables to obtain 2.5D geometry. It limits the study to structures having a minimum volume ratio of 65% corresponding to the stress concentration apparition. However, bone structures generated by nature reach higher porosity rates, which can be achieved by a 3D geometry. Their 3D intersections make possible to reach high levels of porosity while preserving material cohesion. That is why it's necessary to continue the study on a 3D model with a management of the elliptic porosities intersections. It could lead to more efficient structures in terms of weight. New addictive manufacturing technologies allow production of this type of structure. But, to obtain a real weight reduction it is necessary to evacuate the unmelted powder. The generation of 3D internal porosities that can evacuate the unmelted powder can be antagonistic with the generation of mechanically efficient porosities. The development of a 3D metal structure able to be printed by a powder bed technology therefore brings the problem of depowdering that should be solved.

The mechanical assemblies provided by nature are able to sustain a large variety of mechanical load with a high number of cycles. The presented bio-inspired approach opens interesting perspectives for design of mechanical link subject to complex multiaxial loading.

## 4 Conclusion

Additive manufacturing offers new opportunities in the production of porous workpieces with high mechanical performance. This may allow the weight reduction of workpieces. In this paper a novel bio-inspired approach was presented to design openwork workpieces to reduce the weight maintaining an acceptable mechanical strength. The purpose of proposed method was to generate bio-inspired porosities from bones in shape, size and orientation. This method is not based on topological optimization. Two designs were proposed: one inspired from avian species and the second from terrestrial mammalian. Literature highlights that the porosities of bones have an ellipsoid shape and a size depending on the type and size of animal. In this paper a 2.5D pattern was proposed to validate the method usefulness, although the future development to 3D seems even more promising.

The proposed method starts with a finite element analysis of a full dense workpiece in order to extract the stress field. Thanks to an algorithm bio-inspired on the bone structure, ellipsoid porosities were dimensioned and oriented as functions of the principal stress tensor. In order to validate this novel approach, the bio-inspired algorithm was applied on a beam part in 2.5D subjected to a static three-point bending with 65% of density. Three beams were designed and manufactured: two bio-inspired designs (avian and terrestrial) and the last designed by topological optimization method. A finite element analysis was performed on the three beams in order to evaluate the stress field in each beam and predict their behaviors at breaking. Experimental test results had demonstrated the usefulness of the proposed method, especially of the terrestrial mammalian strategy in used experimental conditions. Terrestrial mammalian beam have a flexural stiffness and a failure force respectively increase of 17% and 7.5% compared with the topological optimization. This properties increase was attributed to the homogenization of the stress field thanks to a reduced porosity size.

## 5 Acknowledgments

This work was supported by Aix- Marseille University. The experimental devices were founded by European Community, French Ministry of Research and Education and Aix-Marseille Conurbation Community.